\documentstyle[12pt]{article}

\begin{document}
\setlength{\baselineskip}{0.2in}

\begin{titlepage}
\noindent
\hfill DOE/ER/40427-24-N93

\hfill November 1993

\noindent

\setlength{\oddsidemargin}{0.6in}
\setlength{\evensidemargin}{0.6in}
\setlength{\textwidth}{6.5in}
\setlength{\textheight}{9in}
\setlength{\parskip}{0.05in}
\vspace{36pt}

\vspace{5pt}
\centerline{
\Large{\bf Transverse Quark Distribution in Mesons} }
\vspace{10pt}
\centerline{\Large{\bf - QCD Sum Rule Approach -}}
\par
\par\bigskip
\par\bigskip
\vspace{0.7cm}

\centerline{Su H. Lee\footnote{On leave of absence from Dept. of Physics,
Yonsei Univ., Seoul 120-749, Korea}, T. Hatsuda\footnote{
Present address: Institute of Physics, Univ. of Tsukuba,
Tsukuba, Ibaraki 305, Japan} and G. A. Miller   }

\centerline{\em Department of Physics, FM-15,
University of Washington}
\centerline{\em Seattle, WA 98195, USA}
\vspace{1in}

\centerline{\bf Abstract}
 QCD sum rules are used to compute the first few
moments of  the  mesonic quark momentum.
Transverse,  longitudinal
and mixed transverse-longitudinal  components are examined.
The transverse size of the pion is
 shown to be dictated by the gluon condensate, even though
the mass and the longitudinal distribution
 are dominated by the quark condensate.  The implications
 of our results for color transparency physics and
 finite temperature QCD are discussed.

\end{titlepage}

\newpage
\setcounter{equation}{0}
\renewcommand{\theequation}{\arabic{equation}}

Mesonic wave functions have been the focus of much recent theoretical
activity  regarding
exclusive processes \cite{Sterman92,Dziem87,Huang91} and color transparency
\cite{FmS92}.
In the former case, the asymptotic formula of exclusive process
 can be expressed,  via factorization,
in terms of a hard scattering amplitude $T_H$ and the  quark distribution
function $\phi(\xi,Q^2)$ \cite{Lepage80}
  ($\xi$ is the light-cone momentum fraction of
 quarks and $Q^2$ is the typical external momentum).
However, there have been many suggestions that such formulae are not
valid at presently
accessible energy ranges \cite{Isgur89,Rady91}.  Furthermore, computations
of form factors often employ
a wave function with a spread of $\vec{b}$
(the transverse separation
 between quarks)
 \cite{Sterman92,Jakob93}.
Sterman and collaborators  have computed the
Sudakov effects \cite{Botts89}
for the pion  and proton form factors\cite{Sterman92,Li}.
This involves
``wave function" of the form $e^{S(b,\xi,Q^2)}$.
The Sudakov effect should dominate at high $Q^2$, but it would be
surprising if the $b$-dependence resides only in the function $S$.


The
transverse size dependence plays also a key role in color
transparency physics.
A  Gaussian dependence on the transverse momentum $k^2_\perp$
($\vec{k}_\perp$ is canonically conjugate to
$\vec{b}$) implies that the
effects of color transparency would never be observed, but  a power fall
off does allow such observations \cite{FmS92}.  Furthermore,
QCD lattice calculations \cite{Lattice}  now provide
hadronic wave
functions, and understanding these with analytic techniques
would be useful.

The use of QCD sum rules (QSR) has
 been a particularly useful example of such an
analytic technique.
QSR have long been known to
provide reasonable estimates of hadronic properties \cite{QSR} and
distribution functions needed as inputs to perturbative QCD (PQCD)
 calculations.  In
particular, Chernyak and Zhitnitsky \cite{CZ1,CZ2}
 have determined $\xi^{2n}$
moments of the distribution function $\phi(\xi)$.

The present paper, concerns a new application of
QCD sum rules(QSR): the determination of  the
$\xi^{2l}k^{2n}_\perp$
moments of mesonic wave functions.
We derive  new Borel sum rules for these moments, which makes us
 possible to extract physical picture of the
 quark momentum distributions.
(For an application of a finite energy sum rule, see ref.
\cite{Piv93}.)
The starting point is the Bethe-Salpeter amplitude for the pion
\begin{eqnarray}
\label{def}
i f_\pi p_{\alpha}  \phi_\pi ( z\cdot p,z^2)_\mu
+ z_{\alpha} \chi_{\pi} (z\cdot p,z^2)_\mu & = &
 \langle 0| \bar{d}(z) \gamma_\alpha \gamma^5 e^{ig \int_{-z}^{z} d
\sigma \cdot A} u(-z) |\pi^+(p) \rangle_\mu  \nonumber \\[12pt]
& = &
\sum_n \frac{(i)^n}{n!}
\langle 0| \bar{d}(0) \gamma_\alpha
(iz \cdot \stackrel{\leftrightarrow}{D})^n u(0)   | \pi^+(p) \rangle_\mu  ,
\end{eqnarray}
where $\phi_\pi$ and
$\chi_{\pi}$ represent two different Bethe-Salpeter amplitudes,
$i \stackrel{\leftrightarrow}{D} =i\vec{D}+(i\vec{D})^\dagger$,
$\vec{D}=\vec{\partial}-ig\vec{B}^a \frac{\lambda_a}{2}$,
$\mu$ is the
renormalization scale and  $f_\pi$ is the pion decay constant.
First let  us  introduce
 a light-like four vector $y$ and a transverse vector
$t=(0,\vec{b},0)$ which is perpendicular
 to the hadron momentum $p=(p^0, \vec{0}, p_z)$.
 Then we make a decomposition
 $z=y+t$
so that $z^2=-b^2$, and  $z \cdot p=y \cdot p$.
 Since we
have chosen $z$ to have both longitudinal and transverse components,
 the
matrix element in eq.(1) can be further expanded
in terms of $y$ and $-b^2$. The result is that
\begin{eqnarray}
\label{mnl}
\langle 0| \bar{d}(0) y^\alpha \gamma_\alpha (iz \cdot
\stackrel{\leftrightarrow}{D})^n u(0)   | \pi^+(p)
\rangle_\mu
=i f_\pi
\sum_{l=0}^n (y\cdot p)^{n-l+1} (-b^2)^{\frac{l}{2}}
M_{n-l,l},
\end{eqnarray}
after contracting with $y^\alpha$, and
where the normalization is $M_{0,0}= 1$.  Since $y\cdot z=0$,
the contraction
of  $y^\alpha$ with eq. (1) enables us
to extract  $\phi_{\pi}$ on which we focus our attention in this paper.
$M_{n-l,l}$ are related to the matrix elements with $n-l$ covariant
derivatives in the longitudinal direction and $l$ in the transverse
direction.   Both $n$ and $l$ are even numbers.
The moments $M_{n-l,l}$ can be related to spin $n+1$ and twist $l+
2$ matrix elements by identifying
independent symmetric traceless matrices.
For example,   $-i f_\pi M_{0,2}  p_\mu= \frac{5}{9} i
\langle 0| \bar{d} \gamma_\mu \gamma_5 \sigma_{\alpha \beta}  g G^{\alpha
\beta} u| \pi(p) \rangle$.

The function $\phi_\pi (z\cdot p,z^2)_\mu$ depends on coordinates; the
relation to the momentum space wave function is given by
\begin{eqnarray}
\label{parton}
\phi_\pi (y \cdot p,-b^2)_\mu \equiv \int_{-1}^1 d\xi
\int^{\mu^2} d^2 k_\perp e^{i\xi y\cdot p} e^{-i\vec{k_\perp} \cdot
\vec{b}}
\phi_\pi(\xi,k^2_\perp)_\mu,
\end{eqnarray}
while the light-cone quark distribution function is given by
\begin{eqnarray}
\phi_\pi(\xi)_\mu=\int d^2 k_\perp \theta(\mu^2-k^2_\perp)
\phi_\pi(\xi,k^2_\perp)_\mu.
\end{eqnarray}
Expanding the integrand of eq.(\ref{parton}) as a power series in
$y\cdot p$ and $(b^2)^{1\over 2}$ leads to
\begin{eqnarray}
\label{partonexp}
\phi_\pi (y \cdot p,-b^2)_\mu
= \sum_n \frac{(i)^n}{n!} \sum_{l=0}^n (y\cdot p)^{n-l}
(b^2)^{\frac{l}{2}} \ \
_nC_l \Gamma_l \langle \xi^{n-l} k_\perp^l \rangle_\mu
\end{eqnarray}
where
\begin{eqnarray}
\langle \xi^{n-l} k_\perp^l \rangle_\mu=\int_{-1}^1 \int d^2k_\perp
\xi^{n-l} k_\perp^l \phi_\pi(\xi,k^2_\perp)_\mu,
\end{eqnarray}
$\Gamma_l\equiv \Gamma(1/2+l/2)/(\Gamma(1/2) \Gamma(1+l/2))$ and
$ _nC_l=\frac{n!}{(n-l)!l!}$.
The desired moments are now identified by
using
eqs. (\ref{partonexp}) and eq.(\ref{mnl}) in the contraction of eq. (1)
with $y^\alpha$ and
equating the coefficients of the
terms in the expansion.
 The result is
\begin{eqnarray}
\label{nice}
<\xi^{n-l} k^l_\perp> =\frac{1}{ _nC_l \Gamma_l} M_{n-l,l}(-1)^{l/2}.
\end{eqnarray}

We proceed by using  QCD sum rules to extract  $M_{n-l,l}$ from eq.(\ref{mnl})
and thereby obtain the moments from eq.(\ref{nice}).
Consider the correlation function $T_{n,0}$,
\begin{eqnarray}
\label{cor}
T_{n,0}(y \cdot q,-b^2,q^2) =i \int d^4x e^{iqx} \langle 0|
T[ \bar{d}(x) y^\alpha \gamma_\alpha
\gamma_5 (z \cdot \stackrel{\leftrightarrow}{D}
)^n u(x), \bar{u(0)} y^\alpha \gamma_\alpha
\gamma_5 d(0) ] |0 \rangle.
\end{eqnarray}
The complete sum over intermediate states is approximated by
the lowest mass pion pole term and
a continuum contribution. Then the spectral density
can be expressed as
\begin{eqnarray}
\label{phen}
\frac{1}{\pi} Im T_{n,0}(y \cdot q,-b^2,q^2)=
\sum_{l=0}^n f_\pi^2 (y \cdot q)^{n-l+2}
(-b^2)^{\frac{l}{2}} \left(M_{n-l,l}
\delta(q^2-m_\pi^2)+c(n,l) \theta(q^2-s_n)\right),
\end{eqnarray}
where $c(n,l)$ are coefficients chosen to match the perturbative part
 in the
Operator Product Expansion (OPE)
at $Q^2 \rightarrow \infty$.
We have carried out the OPE for
 $T_{n,0}$  for operators of dimension less than or equal to six:
\begin{eqnarray}
\label{ope}
T_{n,0}(y \cdot q,-b^2,q^2) & = & \sum_{l=0}^n {_nC_l} \Gamma_l
(y \cdot p)^{n-l+2} (b^2)^{l/2} (q^2)^{l/2}
[-C^{pert}_{n,l} {\rm ln}(-q^2)  \nonumber \\[12pt]
& & -(C^1_{n,l}{\rm ln} (-q^2)+C^2_{n,l}) \langle \frac{\alpha}{\pi}G^2
\rangle
\frac{1}{q^4}-
C^4_{n,l}(-1)^{l/2} \langle \sqrt{\alpha}\bar{q}q \rangle^2 \frac{1}{q^6}],
\end{eqnarray}
where
\begin{eqnarray}
C^{pert}_{n,l} & = & \frac{3}{16 \pi^2 } (-1)^{n}
              B(\frac{n-l}{2}+\frac{1}{2},\frac{l}{2}+2) ,
                \nonumber \\[12pt]
C^1_{n,l} & = & \frac{1}{24 }
              B(\frac{n-l}{2}+\frac{1}{2},\frac{l}{2}+1)
              (\frac{l}{2})(\frac{l-2}{2}) \nonumber \\
         & &  [1+ \theta(n-2-l)(n-l)\frac{l+2}{2}
               +\theta(l-2) l \frac{l+3}{2}],
              \nonumber \\[12pt]
C^2_{n,l} & = & \frac{1}{24 }
              B(\frac{n-l}{2}+\frac{1}{2},\frac{l}{2}+1)(l-1)
              \nonumber \\
         & &  [1+ \theta(n-2-l)(n-l)\frac{l+2}{2}
               +\theta(l-2) l \frac{l}{2}\frac{l+1/2}{l-1}],
              \nonumber \\[12pt]
C^{4}_{n,l} & = & + \delta_{l,0} \frac{32 \pi }{81}
              (4n+11) +\delta_{l,2} \frac{2 \pi}{9 _nC_l \Gamma_l} n(n-1),
\end{eqnarray}
with $B(n,l)=\Gamma(n) \Gamma(l)/\Gamma(n+l)$ and $\theta(n)=1$ for
$n \geq 0$.
The OPE for the $\rho$ meson is as above except for the four quark
condensate term $C^{4}_{n,l}$ for which $(4n+11)\rightarrow (4n-7)$.

We may immediately examine the physical consequences of these expressions.
The gluon condensates  arising from the covariant derivative
in eq.(\ref{cor})
give the terms proportional to $\theta$'s in $C^1_{n,l}$ and
 $C^2_{n,l}$.
These terms
together with the other gluon condensate become increasingly important as
$l$ increases.
Conversely, one of the  four quark condensates with hard gluon line
 (Fig. 3 of \cite{CZ1}), which gives the dominant power
correction for the $l=0$ sum rules,  does  not contribute at all
for $l \neq 0$.
This is because the internal gluon line in the
relevant tree graph carries no transverse momentum.
Thus
the power correction is dominated by the gluon condensate and the quark
condensate  plays only a minor role.  This is opposite to the usual
sum rules for
 longitudinal moments ($l=0$) and for the meson masses
where the quark condensate term is essential.
 In
the  constituent quark picture of hadrons, the mass of the
constituent quark is obtained through the dynamical breaking
 of chiral symmetry and the
transverse wave function is determined by the
 confining force  between the constituent
quarks. This picture is consistent
 with our observations about the OPE results.

To go beyond these qualitative aspects we follow
the QSR standard procedure and, equate the phenomenological
side  eq.(\ref{phen}) to the OPE side eq.(\ref{ope}) using the
the standard Borel transformed dispersion relation on $q^2$,
$\frac{1}{\pi}
\int ds e^{-s/M^2} {\rm Im} T_{n,0}(s)={\rm Borel \,\, trans } [{\rm Re}
T_{n,0}]$, and identify the coefficients of the double expansion.
The result is
\begin{eqnarray}
\label{sum}
f_\pi^2  \langle \xi^{n} k_\perp^l \rangle  & = & C^{pert}_{n+l,l}
        (-\frac{d}{d(1/M^2)})^{l/2} M^2
        (1-e^{-s/M^2})  \nonumber \\[12pt]
      & &   + \theta(2-l)C^2_{n+l,l}(-1)^{l/2+1}\frac{1}{(\frac{2-l}{2})!}
         (M^2)^{l/2-1}
          \langle \frac{\alpha}{\pi}G^2 \rangle   \nonumber \\[12pt]
     & & + \theta(l-4)C^1_{n+l,l}(\frac{l-4}{2})! (M^2)^{l/2-1}
           \langle \frac{\alpha}{\pi}G^2\rangle  \nonumber \\[12pt]
     & &    + \theta(2-l)C^4_{n+l,l}
       (-1)^l \frac{1}{(\frac{4-l}{2})!} (M^2)^{l/2-2}
      \langle \sqrt{\alpha}\bar{q}q \rangle^2 \ \ \ .
\end{eqnarray}

The above sum rule reduces to the sum rule for $f_\pi$ \cite{QSR} for $n=l=0$
and to the longitudinal sum rule \cite{CZ1,CZ2} for $l=0$.
We will use the physical value $f_\pi= 133$ MeV and derive
 each moment for
different value of $S_0$.  For the vacuum condensates, we use the
standard values \cite{QSR}:
\begin{eqnarray}
\langle \sqrt{\alpha} \bar{u}u \rangle^2
\simeq 1.83 \cdot 10^{-4}~~ {\rm GeV}^6 , ~~~~~\langle
\frac{\alpha}{\pi} G^2
\rangle \simeq 1.2 \cdot 10^{-2}~~ {\rm GeV}^4
{}.
\end{eqnarray}

The outline of the method used to analyze the sum rule
 is the following:
\begin{enumerate}

\item Find the Borel window,  i.e.,
 find $M^2_{min}$ such that the total power
correction is less than 30 \% of the perturbative part.
This will insure that the neglected higher
dimensional operators cannot be large for $M> M_{min}$.
We then choose $M^2_{max}$ arbitrarily say
 $M^2_{max} =M^2_{min}+0.3 {\rm ~or~ } 0.4 {\rm GeV}^2$.

\item Find the value of $S_0$ which minimizes the dependence on $M^2$.
Then the extracted moments will
not depend too much on the exact choice of $M^2_{max}$.

\item The physical quantity is then obtained from
the average over the Borel window $M^2$.
\end{enumerate}
Chernyak and Zhitnitsky used a similar
 procedure for the case of  $l=0$.

\vskip0.1truein
The form of the Wilson coefficients dictates the characteristic
behavior of the sum rule as a function of the variables $n,l$.
\noindent{\bf Increasing $n$ for fixed $l$} generally enhances the
coefficient of both the gluon and 4-quark condensate  in the OPE.
Then the Borel window (range of $M^2$) appears at higher values of $M^2$ and
the value of $S_0$ increases.  
Note also that the power correction becomes
large as $n$ increases, so that
the sum rule is not reliable for $n >6$.
\noindent{\bf Increasing $l$ for fixed $n$} generally enhances the
coefficient  of the gluon condensate but will reduce that of the
4-quark condensate. In particular,
 there is no 4-quark condensate for $l>2$. This
competing effect tends to keep the Borel window fixed or moves it to
slightly
higher values of $M^2$.  For $l>6$,
the sum rule is again
 sensitive to $S_0$ and one cannot make reliable
estimates.  The value of $S_0$ must be reduced compared to other
moments
to reduce the $M^2$ dependence.

Table 1 shows the result of our analysis.
The moments for $l=0$ correspond to the values obtained in ref.\cite{CZ1,CZ2}.
The main sources of errors in Table 1
are the contribution from higher dimensional operators and the uncertainty
in the exact value
of $S_0$.  The first issue is related to  Mikhailov
and Radyushkin's (MR) \cite{NLC} criticism of ref.\cite{CZ1,CZ2}.
MR
stress  the importance of including non-local condensates \cite{NLC2}.
  Although, there are some ambiguities associated with modeling different
types of
non-local condensates,
 we take the MR results as a guide to estimate
 the uncertainties of our
calculation.  The  Gaussian model of MR leads to
longitudinal moments  about 50 \% smaller than that of ref.\cite{CZ1,CZ2}.
  We expect similar behavior to be true for
sum rules with
$l\neq 0$ and estimate the  uncertainty
associated with unknown higher dimensional operators to be
 $-$50 \% for all
moments.
A second source of errors is the lack of knowledge of $S_0$ which strongly
influences the knowledge of moments with
$l \neq 0$.
For example, increasing the value of $S_0$
to its value for $l=0$, causes a 50\% increase in the
moments. This gives an
upper limit for the $l\neq 0$ moments.  So overall, the estimated errors
for all the moments are $\pm$50 \%.

In the conventional notation, where the separation between quarks is $b$
instead of $2b$,
our $n=0,l=2$ moment implies an average transverse momentum of
($300$MeV)$^2$.
This is a characteristic hadronic scale and is consistent with
a rough estimate ($323$MeV)$^2$  in ref.\cite{CZ2}.

We can use our results to study three features  of the
wave function; approximate
factorization of the longitudinal and transverse directions;
power law behavior of the transverse wave function; and the low temperature
behavior of the transverse wave function.
Let us discuss each one separately.

Here the term ``factorization" refers to the property that
 $\langle \xi^n k_\perp^l \rangle
\sim \langle \xi^n  \rangle \langle k_\perp^l \rangle $.
A recent calculation \cite{jm93}
of color transparency effects in high energy pion-nucleus scattering
assumed this factorization. The results shown in Table 1
are consistent with the factorization property.
The numerical origin of factorization is natural and
can be traced from the expressions provided here.
Further work is in progress to demonstrate that this factorization holds
when including the effects of non-local condensates for $l \neq 0$.

The second point is the $k_\perp $ dependence of
 the transverse wave function.
Once we assume factorization,
the transverse momentum dependence can be examined using the relation
\begin{eqnarray}
\phi(\xi, k_\perp)=\phi(\xi) \psi(k_\perp).
\end{eqnarray}
The knowledge of a finite number of moments is not sufficient
to completely determine
$\psi(k_\perp)$; instead we define a quantity
 $\Delta$ such that
$\Delta\equiv
 [\langle k_\perp^4 \rangle- \langle k_\perp^2
\rangle^2]/ \langle k_\perp^2 \rangle^2$.
We can then determine which of
two popular forms, a Gaussian $\psi_G(k_\perp)=A$exp$(-k_\perp^2/m^2)$
and a power
law $\psi_{PL}(k_\perp)=A\frac{1}{(k_\perp^2+m^2)^2}$,
 has a value of $\Delta$
consistent with the results of Table 1, which is that
  $\Delta \approx 8$.
One may use the $\psi$'s
to calculate $\Delta$; it is first necessary
to introduce a value of $\mu^2$ (see  eqs. (3) and (4))
which is expected to be of the order of 1 GeV$^2$.   For the Gaussian wave
function $\psi_G(k_\perp)$, $\Delta_{\rm G}
\sim 1-O(exp(-\mu^2/m^2))$.  The exponential term
can be estimated by reproducing the value of
$\langle k_\perp^2 \rangle$ shown in Table 1.
For $\mu^2$ larger than 1 GeV$^2$, the
$O(exp(-\mu^2/m^2))$ turns out to be always less than 0.1.
For the power law, we can again   determine $\mu^2$ by reproducing the value
of
$\langle k_\perp^2 \rangle$.  Then $\Delta_{\rm PL}
 =28$ for
$\mu^2 =1$ GeV$^2$ and $\Delta_{\rm PL}=6$ for
$\mu^2 =2$ GeV$^2$ and always larger than 5 for larger values of $\mu^2$.
So the sum rule values
for $\Delta$ are more consistent with a power law type of transverse
wave function even if we take into account the estimated errors.
More generally, the concept of  a large value of $\Delta$
is consistent with the notion of significant fluctuations of the momentum,
a property that favors
the possibility that color transparency would occur \cite{FmS92}.

Another interesting question is the
low temperature generalization of our moment sum rules.   This
generalization is achieved by applying the
low temperature pion gas approximation
to the thermal expectation value of the
hadronic correlation functions.\cite{Ioffe89,HKL93}
The use of soft pion theorems lead to the result  that
to
lowest order in $\epsilon=T^2/6f_\pi^2$,
the axial vector and  vector correlation functions
at finite temperature can be expressed
in terms of a linear combination of vector and axial vector
vacuum correlation functions with temperature dependent
 residues \cite{Ioffe89},
such that the only
effect at finite temperature is a renormalization of $f_\pi^T=
f_\pi(1-\epsilon/2)$.
We find that the  same is  true
 for the transverse moments, i.e.,
\begin{eqnarray}
\label{ft}
M_{0,l}^A(T)=(1-\epsilon)M_{0,l}^A(T=0)+\epsilon M_{0,l}^V(T=0) \nonumber
\\[12pt]
M_{0,l}^V(T)=(1-\epsilon)M_{0,l}^V(T=0)+\epsilon M_{0,l}^A(T=0),
\end{eqnarray}
where the superscripts $V$ and $A$ represent moments for the vector and axial
vector currents.
So the change at finite temperature
can all be accommodated by a change in $f_\pi$ with
no change in the transverse
moments.   This suggests that to lowest order in $T$, the wave functions do
not change when the temperature is increased from zero to a small value.

In summary, we have derived  sum rules for the transverse and
longitudinal moments of the pion (and the $\rho$ meson)
wave function.  These results can be used to understand the role of
the quark wave functions in exclusive processes and at finite temperature.

\vspace{0.5cm}
This work was supported in part by the US Department of Energy under grant
DE-FG06-88ER40427.

\newpage

\centerline{
\begin{tabular}{||c||c|c|c||c|c|c||c|c|c||}   \hline
  & \multicolumn{3}{c||}{$l=0$}  &
\multicolumn{3}{c||}{$l=2$}     &  \multicolumn{3}{c||}{$l=4$}
\\ \cline{2-10}
$ \langle \xi^n k_\perp^l \rangle$  & moments & $M_{min}^2$  & $S_0$ &
moments & $M_{min}^2$  & $S_0$  &
moments & $M_{min}^2$  & $S_0$  \\
&  & (GeV$^2$) & (GeV$^2$) &  (GeV$^2$) & (GeV$^2$) & (GeV$^2$) & (GeV$^4$) &
 (GeV$^2$) & (GeV$^2$)   \\
  \hline \hline
$n=0$  &  1  & 0.6  & 0.9  & 0.36 & 0.8 & 0.5  & 1.17 & 1.2  & 0.4
\\ \hline
$n=2$  &  0.4  & 1.3 & 1.9   & 0.14 & 1.3 & 0.9 &  0.30 & 1.5
& 0.5  \\ \hline
$n=4$  &  0.24 & 1.9 & 2.8  & 0.09 & 1.9 & 1.4 & 0.20 & 2.0
& 0.5  \\ \hline
\end{tabular} }

\vspace{0.3cm}

\centerline{Table 1: Mixed moments of the quark transverse/longitudinal
 momentum distribution.  }


\newpage
\begin{thebibliography}{99}
\bibliographystyle{unsrt}

\newcommand{\btem}{\bibitem}
\newcommand{\PL}{Phys.\,Lett.}
\newcommand{\PTP}{Prog.\,Theor.\,Phys.}
\newcommand{\PR}{Phys.\,Rev.}
\newcommand{\PRL}{Phys.\,Rev.\,Lett.}
\newcommand{\NP}{Nucl.\,Phys.}
%
\setlength{\itemsep}{0.0in}

\btem{Sterman92} H-n Li and G. Sterman,
Nucl.\,Phys.\, {\bf B381} (1992) 129;

\btem{Dziem87} Z. Dziembowski and L. Mankiewicz,
\PRL\  {\bf 58} (1987) 2175.

\btem{Huang91} T. Huang. Q. X. Shen,
Z.\,Phys.\, {\bf C50} (1991) 139.

\btem{FmS92} L. Frankfurt, G. A. Miller and M. Strikman,
Comments Nucl.\, Part.\, Phys.\,{\bf 21} (1992) 1; Nucl.
Phys. {\bf A555} (1993) 752.

\btem{Lepage80} G. P. Lepage and  S. J. Brodsky,
\PR\  {\bf D22} (1980) 2157.

\btem{Isgur89} N. Isgur and C.H. Llewellyn Smith,
Nucl.\,Phys.\ {\bf B317} (1989) 526.

\btem{Rady91} A. P. Bakulev and A. V. Radyushkin,
\PL\  {\bf B271} (1991) 223.

\btem{Jakob93} R. Jakob and P. Kroll,
\PL\ {\bf B315} (1993) 463.

\btem{Botts89} J. Botts and G. Sterman,
Nucl.\,Phys.\, {\bf B325} (1989) 685.

\btem{Li}
H-n Li,  (SUNY, Stony Brook), \PR\
{\bf D48}  (1993) 4243.

\btem{Lattice} D. Daniel et.al.,
\PR\  {\bf D46} (1992) 3130.
M.C. Chu, M. Lissia and J.W. Negele, \NP\  {\bf B360} (1991) 31.
 R. Gupta, D. Daniel and J. Grandy,
`` Bethe-Salpeter Amplitudes and Density Correlations for
 Mesons with Wilson Fermions", LA UR-93-1051.

\btem{QSR} M.\,A.\,Shifman, A.\,I.\,Vainshtein and V.\,I.\,Zakharov,
Nucl.\,Phys.\ {\bf B147} (1979) 385, 448.
L.\,J.\,Reinders, H.\,Rubinstein and S.\,Yazaki, Phys.\,Rep.\ {\bf
127} (1985).

\btem{CZ1} V. L. Chernyak and Z. R. Zhitnitsky,
Nucl.\,Phys.\ {\bf B201} (1982) 492.

\btem{CZ2} V. L. Chernyak and Z. R. Zhitnitsky,  \, Phys.\,Rep.\, {\bf 112}
(1984)173.

\btem{Piv93} A. A. Pivovarov,
Nucl.\,Phys.\,{\bf B396} (1993) 119.

\btem{NLC} S. V. Mikhailov and A. V. Radyushkin, Phys. Rev. {\bf D 45}
 (1992) 1754.

\btem{NLC2} E. V. Shuryak, \NP\, {\bf B203}
 (1982) 116; D. Gromes, \PL\, {\bf B115} (1982) 482.

\btem{jm93} L. Frankfurt, G.A. Miller and M. Strikman, Phys.Lett.
   {\bf B304} (1993) 1.

\btem{Ioffe89} M. Dey, V. I. Eletsky and B.L. Ioffe,
\PL\, {\bf B252} (1990) 620.

\btem{HKL93} T. Hatsuda, Y. Koike and S. H. Lee,
Nucl.\,Phys.\, {\bf B394} (1993) 221.



\end{thebibliography}
\end{document}